\documentclass[aps,twocolumn,superscriptaddress,showpacs,amsmath,amssymb,longbibliography]{revtex4-1}
\usepackage{epsfig}
\usepackage{subfigure}
\usepackage{amssymb,amsmath}
\usepackage{graphicx}
\usepackage{epstopdf}
\usepackage{epsfig}
\usepackage{color}
\usepackage{amsfonts}
\usepackage{amscd}
\usepackage{float}
\usepackage[colorlinks,urlcolor=blue,linkcolor=blue,citecolor=blue]{hyperref}
\usepackage{bbm}
\usepackage{mathrsfs}
\usepackage{mathtools}

\emergencystretch=\maxdimen
\DeclareMathSizes{10}{10}{7}{5}

\begin{document}

\title{Coherent-cluster-state generation in networks of degenerate optical parametric oscillators}

\author{Zheng-Yang Zhou}
\affiliation{\mbox{Key Laboratory of Optical Field Manipulation of Zhejiang Province, Department of Physics, ZSTU, Hangzhou 310018, China}}
\affiliation{Theoretical Quantum Physics Laboratory, Cluster for Pioneering Research, RIKEN, Wakoshi, Saitama 351-0198, Japan}
\affiliation{Beijing Computational Science Research Center, Beijing
100094, China}
\author{Clemens Gneiting}
\affiliation{Theoretical Quantum Physics Laboratory, Cluster for Pioneering Research, RIKEN, Wakoshi, Saitama 351-0198, Japan}
\affiliation{Center for Quantum Computin, RIKEN, Wakoshi, Saitama 351-0198, Japan}
\author{J. Q. You}
\altaffiliation[jqyou@zju.edu.cn]{}
\affiliation{\mbox{Interdisciplinary Center of Quantum Information, State Key Laboratory of Modern Optical Instrumentation,} Zhejiang Province Key Laboratory of Quantum Technology and Device, School of Physics, Zhejiang University, Hangzhou 310027, China}
\affiliation{Beijing Computational Science Research Center, Beijing
100094, China}
\author{Franco Nori}
\altaffiliation[fnori@riken.jp]{}
\affiliation{Theoretical Quantum Physics Laboratory, Cluster for Pioneering Research, RIKEN, Wakoshi, Saitama 351-0198, Japan}
\affiliation{Center for Quantum Computing, RIKEN, Wakoshi, Saitama 351-0198, Japan}
\affiliation{Physics Department, The University of Michigan, Ann Arbor, Michigan 48109-1040, USA}

\date{\today}

\begin{abstract}
Cluster states are versatile quantum resources and an essential building block for measurement-based quantum computing. The possibility to generate cluster states in specific systems may thus serve as an indicator regarding if and to what extent these systems can be harnessed for quantum technologies and quantum information processing in particular. Here, we apply this analysis to networks of degenerate optical parametric oscillators (DOPOs), also called coherent Ising machines (CIMs). CIMs are distinguished by their highly flexible coupling capabilities, which makes it possible to use them, e.g., to emulate large spin systems. As CIMs typically operate with coherent states (and superpositions thereof), it is natural to consider cluster states formed by superpositions of coherent states, i.e., coherent cluster states. As we show, such coherent cluster states can, under ideal conditions, be generated in DOPO networks with the help of beam splitters and classical pumps. Our subsequent numerical analysis provides the minimum requirements for the generation of coherent cluster states under realistic conditions. Moreover, we discuss how nonequilibrium pumps can improve the generation of coherent cluster states. In order to assess the quality of the cluster-state generation, we map the generated states to an effective spin space using modular variables, which allows us to apply entanglement criteria tailored for spin-based cluster states.
\end{abstract}

\maketitle

%
%

\section{Introduction}
Cluster states~\cite{10.1103/PhysRevLett.86.910} constitute a generic class of highly entangled quantum states, which can serve as the starting point for subsequent entirely measurement-driven universal quantum information processing~\cite{10.1103/PhysRevLett.86.5188, 10.1103/PhysRevLett.95.010501,10.1103/PhysRevLett.97.110501, 10.1103/PhysRevLett.97.120501,10.1126/science.1142892,10.1103/PhysRevLett.102.100501,10.1103/PhysRevA.81.052332, 10.1103/PhysRevLett.112.120504, 10.1103/PhysRevLett.112.140505,10.1103/PhysRevX.5.041007, 10.1103/PhysRevX.7.041023,10.1103/PhysRevA.105.042610}. In addition, cluster states are useful for quantum sensing~\cite{10.1103/PhysRevA.79.022103,10.1103/PhysRevA.102.052601} and robust to specific noise sources~\cite{10.1103/PhysRevLett.92.180403,10.1103/PhysRevA.102.032614}. While cluster states were originally proposed for spin systems~\cite{10.1103/PhysRevLett.86.910,10.1103/PhysRevLett.97.230501,10.1103/PhysRevA.75.052319}, they are now experimentally accessible~\cite{10.1038/nphys507, 10.1103/PhysRevLett.98.070502,10.1103/PhysRevA.78.012301,10.1038/nphys1777,10.1103/PhysRevA.82.032305,10.1038/nphoton.2013.287,10.1038/s41567-018-0347-x, 10.1126/science.aah4758,10.1126/science.aay2645,10.1002/qute.202200031} on a variety of different platforms. In particular, optical cluster states ~\cite{10.1126/science.1142892,10.1038/nphys507,10.1103/PhysRevLett.98.070502, 10.1126/science.aay2645, 10.1103/PhysRevA.73.032318, 10.1103/PhysRevA.76.032321, 10.1103/PhysRevA.90.032325, 10.1103/PhysRevA.101.043832} promise high coupling flexibility and good scaling properties. As bosonic modes are described by continuous variables, optical cluster states are usually encoded in a subspace of the total bosonic space. Typical choices are a few-photon Fock space~\cite{10.1038/nphys507}, a GKP-code space~\cite{10.1103/PhysRevA.64.012310, 10.1103/PhysRevLett.112.120504}, and a coherent-state basis~\cite{10.1016/j.physleta.2008.02.009, 10.1016/j.physleta.2009.05.059, 10.1088/0953-4075/42/8/085502}.

The coherent Ising machine (CIM)~\cite{10.1364/OE.19.018091, 10.1103/PhysRevA.88.063853, 10.1038/nphoton.2014.249,10.1126/science.aah5178, 10.1126/science.aah4243,10.1038/nphoton.2016.68, 10.1038/s41534-017-0048-9, 10.1103/PhysRevA.96.053834,arXiv:2204.00276v1,10.1038/s41567-021-01492-w,10.48550/arXiv.2212.02598} is optical hardware with the simulation of many-body spin systems being a target application. Similar to platforms for the generation of optical cluster states, the CIM describes a network composed of degenerate optical parametric oscillators (DOPOs)~\cite{10.1080/713820226,10.1103/PhysRevLett.59.2555,10.1103/PhysRevA.43.6194,10.1103/PhysRevLett.86.2770}. Such DOPO networks have, by virtue of their highly flexible couplings among the modes, already been successfully applied in the semiclassical regime, e.g., to solve combinatorial optimization problems~\cite{10.1038/nphoton.2014.249,10.1126/science.aah5178, 10.1126/science.aah4243,10.1038/nphoton.2016.68, 10.1038/s41534-017-0048-9}.

In this article, we address the question of if and to what extent the CIM can be operated and utilized in the quantum regime. To this end, we investigate the possibility to generate cluster states with the CIM. In contrast to, and beyond, the production of entangled cat states~\cite{10.1103/PhysRevA.104.013715}, the successful generation of cluster states would provide clear evidence of the presence of exploitable and processable quantum resources, with the prospect of universal quantum computation looming. As these machines generically operate with coherent states, we focus here on coherent cluster states.

The Ising interaction, realizable by design in the CIM, can serve as a natural basis for the generation of cluster states~\cite{10.1103/PhysRevLett.86.910, 10.1103/PhysRevA.71.034308}. Moreover, CIMs associate coherent states with different phases to different spin orientations, which suggests a straightforward translation of spin cluster states to coherent cluster states. However, we must expect that the cluster-state generation is impaired by the unavoidable presence of single-photon loss and the related trade-off to implement in-principle adiabatic evolutions in finite time. Here, we analyze the minimum requirements such that the generation of cluster states remains successful under these realistic conditions.

Following the same strategy as for entangled cat states~\cite{10.1103/PhysRevA.104.013715}, we employ tailored entanglement criteria in order to assess the successful generation of coherent cluster states. This allows us to certify the presence of cluster states even if the states generated under realistic conditions are imperfect and mixed. While the entanglement criteria designed for cluster states typically assume spin-like Hilbert spaces~\cite{10.1103/RevModPhys.80.517, 10.1103/PhysRevA.69.052327, 10.1103/PhysRevA.69.062311, 10.1103/PhysRevLett.92.087902, 10.1103/PhysRevLett.95.210502, 10.1103/PhysRevA.72.022340, 10.1103/PhysRevLett.98.180502, 10.1016/j.physrep.2009.02.004} and thus cannot be directly applied to coherent cluster states, we use modular variables~\cite{10.1007/BF00670008} in order to map coherent states and cat-like states to effective spin states~\cite{10.1103/PhysRevA.81.052327, 10.1103/PhysRevA.94.022325, 10.1103/PhysRevX.8.021001, 10.1103/PhysRevLett.125.040501, 10.1103/PhysRevA.104.012431}. In these effective spin spaces, we can apply the entanglement criteria for spin cluster states in order to certify the presence of coherent cluster states. Indeed, modular variables have already been successfully deployed for the detection of entanglement in non-Gaussian states~\cite{10.1103/PhysRevLett.106.210501, 10.1103/PhysRevA.86.032332, 10.1103/PhysRevA.88.013610, 10.1103/PhysRevA.90.022115,10.1103/PhysRevA.92.022308}.

This article is structured as follows: We first introduce the theoretical model in the ideal case. Next, we derive the conditions for the generation of coherent cluster states in DOPOs in the presence of detrimental effects, in particular, single-photon loss and nonadiabatic evolution. To this end, the generated candidate cluster states are mapped onto effective spin spaces using modular variables and evaluated with an entanglement criterion for spin cluster states. After studying and optimizing the influences of different parameters, we finally discuss the potential benefits of nonequilibrium pumps.

\section{Theoretical model}
\subsection{Structure of DOPO networks in CIMs}
\begin{figure}[t]
\center
\includegraphics[width=3.4in]{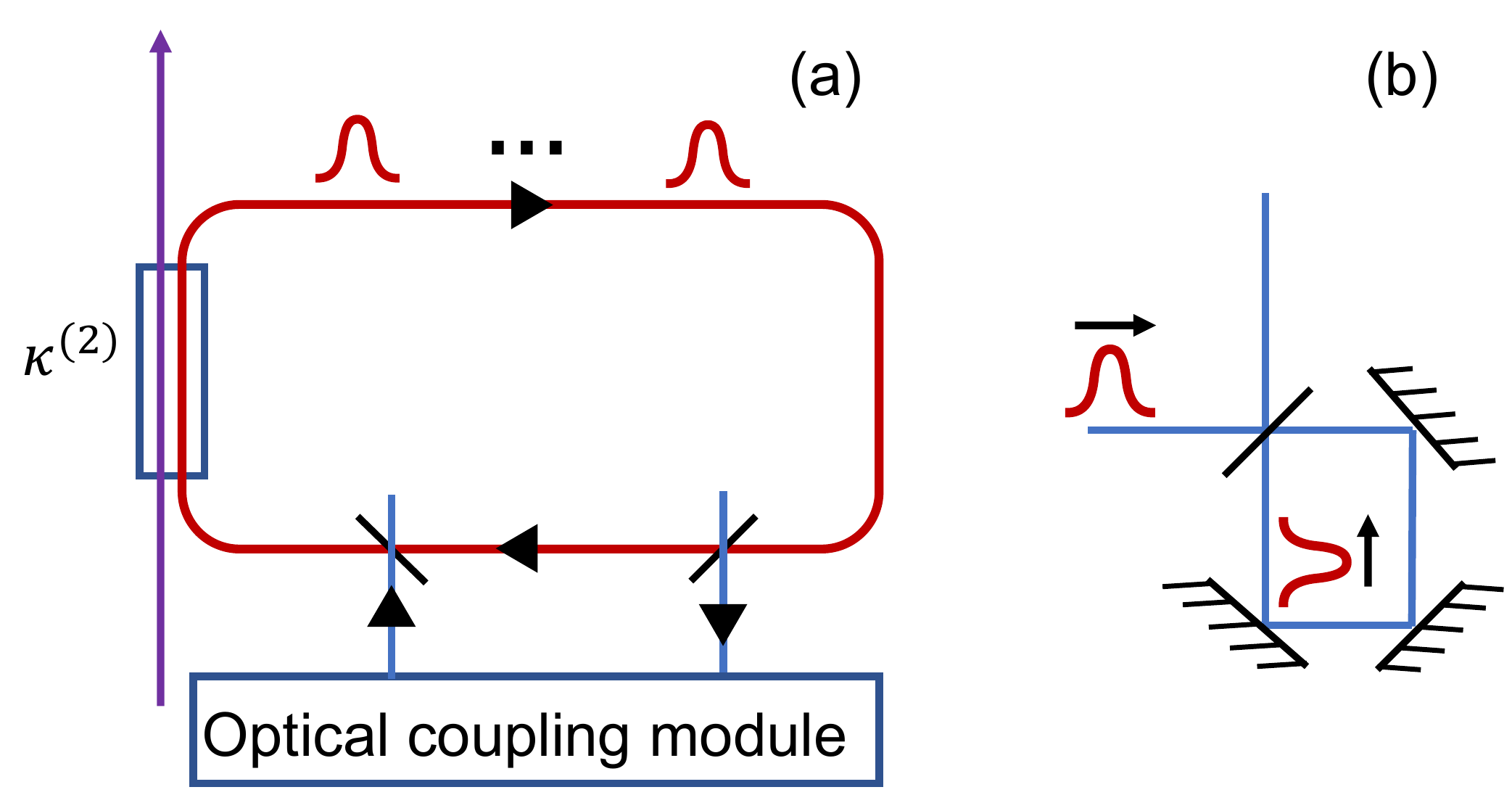}
\caption{(a) Illustration of the structure of a CIM. Optical pulses, which are pumped in the nonlinear crystal, travel in the optical fiber loop. The coupling between pulses is realized by optical delay lines or measurement feedback. (b) Illustration of the beam-splitter interaction between neighboring DOPO modes, which implements the Ising interaction (\ref{coherentising}).}\label{fig1}
\end{figure}
To explore the extension of cluster states to DOPO systems, let us begin by reviewing the theoretical model of DOPO networks in a CIM. A DOPO system consists of a nonlinear component and a cavity~\cite{10.1080/713820226,10.1103/PhysRevLett.59.2555,10.1103/PhysRevA.43.6194,10.1103/PhysRevLett.86.2770}. In particular, DOPOs within a CIM are implemented using a nonlinear waveguide and a fiber cavity, as illustrated in Fig.~{\ref{fig1}}(a). The optical modes correspond to signal pulses traveling in the optical fiber loop. When a signal pulse interacts with the nonlinear waveguide, it couples to the pump pulse through the following Hamiltonian (we set $\hbar=1$):
\begin{eqnarray}\label{nonlinearhamiltonian}
H_{\rm nl}=g_{\rm nl}(a_{\rm s}^2a_{\rm p}^{\dag}+{a_{\rm s}^{\dag}}^2a_{\rm p}),
\end{eqnarray}
where $a_{\rm s}$ and $a_{\rm p}$ correspond to the annihilation operator of the signal mode and the annihilation operator of the pump mode, respectively. If the length of the nonlinear wave guide is short or the loss for the pump mode is strong, the pump mode in Eq.~(\ref{nonlinearhamiltonian}) can be adiabatically eliminated. Such an adiabatic model can be described by a two-photon pump
\begin{eqnarray}\label{twophotonpump}
H=-iS[(a_{\rm s}^{\dag})^2-(a_{\rm s})^2],
\end{eqnarray}
and two-photon loss~\cite{10.1088/0305-4470/11/2/018,10.1088/0305-4470/11/2/018,10.1103/PhysRevA.48.1582,10.1103/PhysRevA.49.2785,10.1088/0305-4470/30/9/008,10.1103/PhysRevA.55.3842,10.1088/1367-2630/16/4/045014,10.1103/PhysRevA.90.033831},
\begin{eqnarray}\label{twophotoloss}
\mathcal{L}_{\rm tp}(\rho)&=&\frac{\Gamma_{\rm d}}{2}[2a_{\rm s}a_{\rm s}\rho(t)a_{\rm s}^{\dag}a_{\rm s}^{\dag}-\{a_{\rm s}^{\dag}a_{\rm s}^{\dag}a_{\rm s}a_{\rm s},\rho(t)\}],\nonumber\\
\end{eqnarray}
where $\{\bullet,\bullet\}$ denotes the anti-commutator, and $\rho$ is the density matrix of the signal mode. Note that we omit the subscript ``s" in the following part. Such a DOPO has a steady-state space formed by the superposition of two coherent states with different phases (i.e., a cat state),
\begin{eqnarray}{\label{darkspace}}
|\Psi(t\rightarrow\infty)\rangle=C_{+}|\alpha\rangle+C_{-}|-\alpha\rangle.
\end{eqnarray}
The complex amplitude $\alpha$ of the coherent states is given by $\alpha=i\sqrt{2S\Gamma_{\rm d}}$. In the presence of single-photon loss,
\begin{eqnarray}\label{singlephotoloss}
\mathcal{L}_{\rm s}(\rho)&=&\frac{\Gamma_{\rm d}}{2}[2a\rho(t)a^{\dag}-\{a^{\dag}a,\rho(t)\}],\nonumber\\
\end{eqnarray}
the steady states are usually mixed states:
\begin{eqnarray}{\label{darkspacemixed}}
\rho(t\rightarrow\infty)=P_{+}|\alpha\rangle\langle\alpha|+P_{-}|-\alpha\rangle\langle-\alpha|.
\end{eqnarray}
This subspace can be used to emulate a spin,
\begin{eqnarray}\label{coherentspin}
|\alpha\rangle\longleftrightarrow|\uparrow\rangle,~~|-\alpha\rangle\longleftrightarrow|\downarrow\rangle,
\end{eqnarray}
for large amplitudes, $|\alpha|\gg1$ or $|\langle\alpha|-\alpha\rangle|\sim0$. Note that, for general values of $\alpha$, two cat states with different parities can alternatively be used to express spins.

The coupling between different effective spins in a CIM is realized by an optical coupling module as illustrated in Fig.~\ref{fig1}(a). Two common design principles for the optical coupling modular are the optical delay-line structure~\cite{10.1038/nphoton.2014.249} and the measurement-feedback structure~\cite{10.1126/science.aah5178}. The optical delay-line coupling can be described by a collective loss, e.g.,
\begin{eqnarray}\label{collectiveloss}
\mathcal{L}_{\rm c}(\rho)&=&\Gamma_{\rm c}(a_i+a_j)\rho(t)(a_i^\dag+a_j^{\dag})\nonumber\\
&&-\frac{\Gamma_{\rm c}}{2}\{(a_i^{\dag}+a_j^{\dag})(a_i+a_j),\rho(t)\},
\end{eqnarray}
where $i$ and $j$ correspond to two different DOPO modes. In contrast, the measurement-feedback coupling can be expressed as classical pumps on different modes:
\begin{eqnarray}\label{mfc}
H_{\rm MF}&=&\sum_i\Omega_i(X_1,\ldots)(a_i+a_i^\dag),
\end{eqnarray}
where $X_j$ is the result of a measurement (usually a position measurement) on the $j$th DOPO mode, and $\Omega_i(X_1,\ldots)$ is the measurement-based pump intensity on the $i$th mode.

Note that the current CIM design does not contain direct coupling, i.e., $(a_ia_j^{\dag}+a_j^{\dag}a_i)$, but such coupling can be realized by a small loop structure, as illustrated in Fig.~\ref{fig1}(b).

\subsection{Coherent cluster states in DOPOs}

Cluster states~\cite{10.1103/PhysRevLett.86.910} are a class of highly entangled states that can be used for measurement-based quantum information processing. A typical (one-dimensional) 1-D spin cluster state with $N$ sites assumes the following form:
\begin{eqnarray}\label{1-Dscluster}
|{\rm Cluster}\rangle^N_{\rm s}&=&\frac{1}{2^{N/2}}\bigotimes_{i=1}^N(\sigma_{z}^{(i+1)}|\uparrow_{i}\rangle+|\downarrow_{i}\rangle).
\end{eqnarray}
By design, these states satisfy a recursive decomposition property under local measurements,
\begin{eqnarray}\label{1-Dsclusterro}
\langle\downarrow_{k}|{\rm Cluster}\rangle^N_{\rm s}&=&\frac{1}{2^{N/2}}\bigotimes_{i=1}^{(k-1)}(\sigma_{z}^{(i+1)}|\uparrow\rangle_{i}+|\downarrow\rangle_{i})\nonumber\\
&&\bigotimes_{i=k+1}^N(\sigma_{z}^{(i+1)}|\uparrow\rangle_{i}+|\downarrow\rangle_{i})\nonumber\\
&=&\frac{1}{\sqrt{2}}|{\rm Cluster}\rangle^{k-1}_{\rm s}\otimes|{\rm Cluster}\rangle^{N-k}_{\rm s} . \nonumber\\
\end{eqnarray}
In particular, this implies that cluster states can, in general, remain entangled after local measurements.

Based on these optical spins in Eq.~(\ref{coherentspin}), it is now straightforward to formulate the corresponding coherent cluster states,
\begin{eqnarray}\label{coherentcluster}
|{\rm Cluster}\rangle^N_{\rm co}&=&\frac{1}{2^{N/2}+\varepsilon}\bigotimes_{i=1}^{N}\left(\frac{a_{i+1}}{\alpha}|\alpha_i\rangle+|-\alpha_i\rangle\right) ,
\end{eqnarray}
where $\varepsilon$ accounts for the nonvanishing overlap between $|\alpha\rangle$, and $|-\alpha\rangle$, and the subscript ``co" stands for coherent cluster state. Specifically, the coherent cluster state for the two-mode case, which we will consider in more detail below, takes the following form:
\begin{eqnarray}\label{coherentcluster2m}
|{\rm Cluster}\rangle^2_{\rm co}&=&\frac{1}{2+\varepsilon'}(|\alpha\rangle\otimes|\alpha\rangle+|\alpha\rangle\otimes|-\alpha\rangle\nonumber\\
&&+|-\alpha\rangle\otimes|\alpha\rangle-|-\alpha\rangle\otimes|-\alpha\rangle).
\end{eqnarray}
\subsection{Generation of coherent cluster states}

Spin cluster states can be generated by utilizing the Ising interaction between different spins~\cite{10.1103/PhysRevLett.86.910,10.1103/PhysRevA.71.034308}. According to this method, $N$ spins or qubits are first prepared in the ground state $|0_1\rangle\otimes|0_2\rangle \ldots\otimes|0_N\rangle$. A rotation $U_{x}=\exp(i\sum_{n=1}^N{\sigma^{(n)}_x}\pi/2)$ is then applied to all the qubits to transform the state into
\begin{eqnarray}\label{spinplus}
|+\rangle\equiv\frac{1}{2^{N/2}}(|\uparrow\rangle+|\downarrow\rangle)\otimes(|\uparrow\rangle+|\downarrow\rangle)\ldots.
\end{eqnarray}
Finally, the Ising interaction is applied, described by the Hamiltonian
\begin{eqnarray}\label{sising}
H_{\rm s-int}=g\sum_{<i,j>}\frac{1+\sigma_{z}^{(i)}}{2}\frac{1-\sigma_{z}^{(j)}}{2}.
\end{eqnarray}
Note that the subscript ``s-int" denotes the spin interaction. A typical choice of $i$ and $j$ is nearest neighbors, i.e., $j=i+1$, in which case the following cluster state can be generated:
\begin{eqnarray}\label{1-Dsclusterge}
|{\rm Cluster}\rangle_{\rm s}&=&\exp({-i\frac{\pi}{g}H_{\rm s-int}})|+\rangle\nonumber\\
&=&\frac{1}{2^{N/2}}\bigotimes_{i=1}^N(\sigma_{z}^{(i+1)}|\uparrow\rangle+|\downarrow\rangle).
\end{eqnarray}

A similar state generation protocol can be realized with DOPOs. Recall that a single DOPO initialized in the vacuum state assumes a cat state as the steady state. Correspondingly, $N$ independent DOPOs assume a steady state similar to $|+\rangle$ in Eq.~(\ref{spinplus}):
\begin{eqnarray}\label{catplus}
|+_{\rm cat}\rangle\equiv\frac{1}{(2+\varepsilon_{\rm mc})^{N/2}}\bigotimes_{i=1}^N(|\alpha\rangle+|-\alpha\rangle).
\end{eqnarray}
Moreover, the nearest-neighbor Ising interaction Hamiltonian (\ref{sising}) can be realized by beam splitters and classical pumps in the adiabatic limit~\cite{10.1088/1367-2630/16/4/045014},
\begin{eqnarray}
\sigma_zdt&\sim& \frac{1}{2\alpha}a^{\dag}dt+{\rm H.c.},\nonumber\\
\sigma_z^{(i)}\sigma_z^{(j)}dt&\sim& \frac{1}{2\alpha\alpha^*}(a_ia_j^{\dag}+a_i^{\dag}a_j)dt,
\end{eqnarray}
with the complex amplitude $\alpha$ of the optical spin~(\ref{coherentspin}). The classical pump term,
$$1/(2\alpha)(a^{\dag}+a),$$
can be realized by the feedback (\ref{mfc}) in the current CIM design framework, while the beam splitter coupling, $$1/(2\alpha\alpha^*)(a_ia_j^{\dag}+a_i^{\dag}a_j),$$
is distinct from both above mentioned coupling methods in the CIM. Such a coupling can be realized by a loop structure as illustrated in Fig.~\ref{fig1}(b). This structure can be either directly inserted into the main loop in Fig.~\ref{fig1}(a) or attached as a parallel pass way through optical switchers~\cite{10.1038/s41566-022-01044-5}.

Accordingly, the Ising interaction Hamiltonian can be implemented through
\begin{eqnarray}\label{coherentising}
H_{\rm co-int}=\frac{g_{\rm c}}{8\alpha}\sum_{i}\left[\left(a_i^{\dag}-a_{i+1}^{\dag}+\frac{1}{\alpha^*}a_{i}a_{i+1}^{\dag}\right)+h.c.\right].\nonumber\\
\end{eqnarray}
Here, the subscript ``co-int" denotes the coherent-state interaction. Coherent cluster states are now generated as
\begin{eqnarray}
\rho_{\rm Cluster}&=&\int_0^{\frac{\pi}{g_{\rm c}}}ds\{-i[H_{\rm co-int}+H,\rho(s)]+\mathcal{L}_{\rm d}(\rho(s))\}+\rho(0)\nonumber\\
\end{eqnarray}
with
\begin{eqnarray}
\rho(0)\equiv|+_{\rm cat}\rangle\langle+_{\rm cat}|.\nonumber
\end{eqnarray}
Note that the effective coherent Ising interaction (\ref{coherentising}) requires the evolution to be adiabatic, $g_{\rm c}\ll \Gamma_{\rm d}$. Therefore, ideal coherent cluster states cannot be generated in practice, due to the finite evolution time and the presence of single-photon loss. To characterize the generated imperfect cluster states, we use an entanglement criterion, which we introduce next.

\subsection{Cluster-state entanglement verification with modular variables}

The entanglement of spin cluster states can be detected with the respective stabilizer entanglement criterion for cluster states~\cite{10.1103/PhysRevA.69.052327, 10.1103/PhysRevA.72.022340}. Specifically, a state of $N$ spins exhibits cluster-state entanglement if the entanglement witness $\mathcal{W}$ satisfies
\begin{eqnarray}\label{spinclusterentanglementcriterion}
\mathcal{W}&\equiv&\sum_{n}\langle\sigma_z^{(n-1)}\sigma_x^{(n)}\sigma_z^{(n+1)}\rangle^2\geq\frac{N}{2}, \\
\sigma_z^{(-1)}&=&\sigma_z^{(N+1)}=I. \nonumber
\end{eqnarray}
By construction, an ideal 1-D cluster state (\ref{1-Dscluster}) with $N$ spins satisfies the entanglement witness (\ref{spinclusterentanglementcriterion}). More generally, imperfect, mixed states still satisfy the witness as long as they maintain essential characteristics of cluster states. We thus can take the entanglement witness (\ref{spinclusterentanglementcriterion}) in order to assess the presence of cluster states even if the actual generated states are imperfect and mixed. In contrast to other proximity measures, such as the fidelity, satisfaction of the entanglement witness here provides a clear division between ``cluster state-like'' and not.

Due to the similarity between the spin cluster states (\ref{1-Dscluster}) and the coherent cluster states (\ref{coherentcluster}), it is natural to extend the spin-based criterion (\ref{spinclusterentanglementcriterion}) to DOPO systems. However, the respective entanglement criterion cannot be based on a naive mapping according to Eq.~(\ref{coherentspin}), as the latter is not applicable to general continuous-variable states. To obtain a generally valid procedure, we use the fact that effective Pauli operators for continuous variables can be formulated as~\cite{10.1007/BF00670008,10.1007/BF00672451}

\begin{eqnarray}\label{opticalPaulimatrices}
\bar{\sigma}_x&\equiv&{\rm cos}(\hat{p}l)-i{\rm sin}(\hat{p}l)\frac{{\rm sin}(\pi\hat{x}/l)}{|{\rm sin}(\pi\hat{x}/l)|},\nonumber\\
\bar{\sigma}_y&\equiv&{\rm sin}(\hat{p}l)+i{\rm cos}(\hat{p}l)\frac{{\rm sin}(\pi\hat{x}/l)}{|{\rm sin}(\pi\hat{x}/l)|},\nonumber\\
\bar{\sigma}_z&\equiv&\frac{{\rm sin}(\pi\hat{x}/l)}{|{\rm sin}(\pi\hat{x}/l)|} .
\end{eqnarray}
By construction, these operators have the same commutation relation as the Pauli matrices $\sigma_x$, $\sigma_y$, and $\sigma_z$.

Based on this correspondence, we can in principle evaluate the stabilizer entanglement criterion~(\ref{spinclusterentanglementcriterion}) with continuous-variable states directly, by substituting the operators ~(\ref{opticalPaulimatrices}). It is more instructive and practical, however, to map the continuous-variable state to an effective spin state first, and then to evaluate~(\ref{spinclusterentanglementcriterion}). Specifically, we identify a decomposition of the continuous variable such that the effective Pauli operators~(\ref{opticalPaulimatrices}) satisfy
\begin{eqnarray}\label{espindecomposed}
\bar{\sigma}_x&=&\sigma_x\otimes I_{\rm anci},\nonumber\\
\bar{\sigma}_y&=&\sigma_y\otimes I_{\rm anci},\nonumber\\
\bar{\sigma}_z&=&\sigma_z\otimes I_{\rm anci},
\end{eqnarray}
\begin{figure}[t]
\center
\includegraphics[width=3.4in]{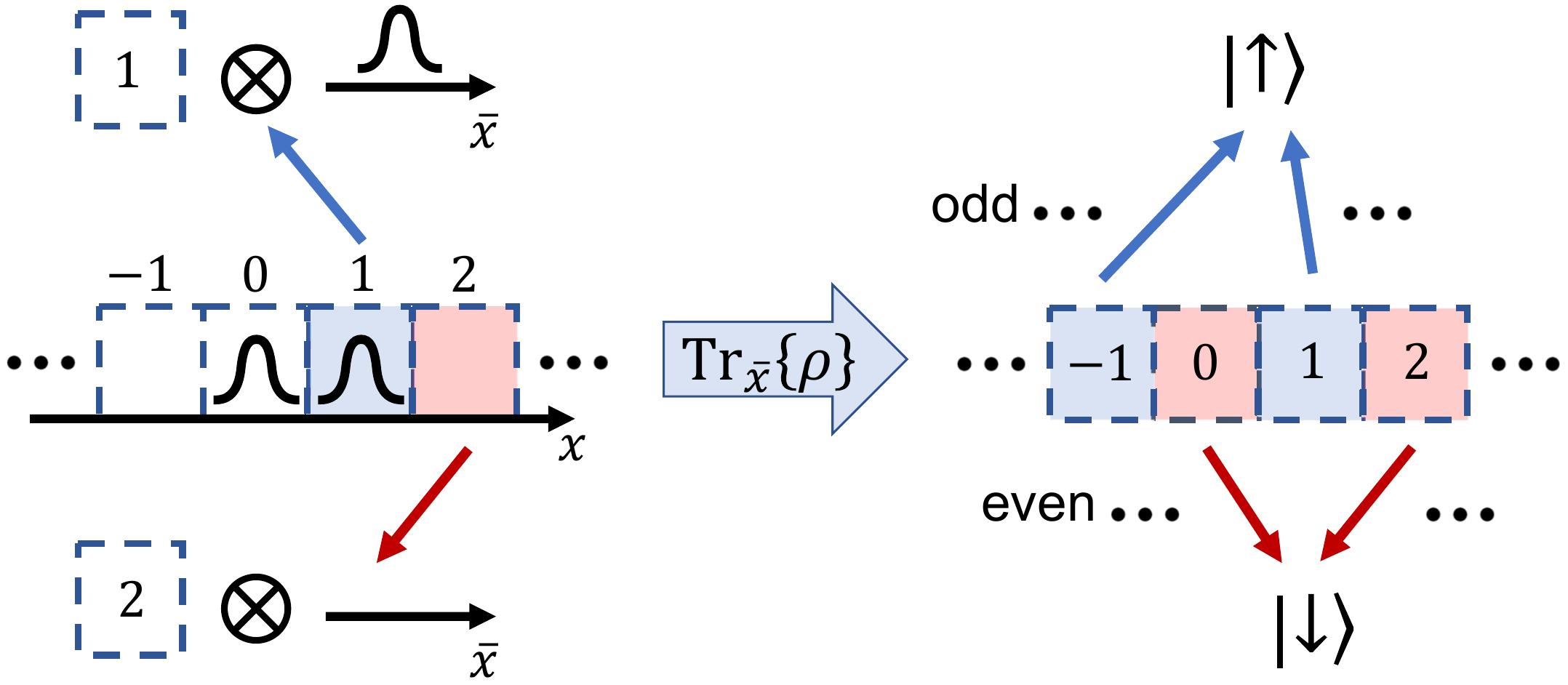}
\caption{Mapping a bosonic mode (i.e., a continuous-variable system) onto an effective spin using modular variables. The continuous variable is split into integer parts (dashed boxes) and modular parts (wave functions in different boxes). By tracing out the modular parts, the continuous variable is then mapped to a discrete-level chain (dashed boxes). This chain is then further restructured as a chain of cells with two internal levels. The latter internal space constitutes an effective spin.}\label{fig2}
\end{figure}
that is, after tracing out the ancilla space, we preserve the Pauli algebra in the effective spin space.

To determine the ancillary space, we consider the position basis and express the position eigenstates using modular variables~\cite{10.1007/BF00670008,10.1103/PhysRevA.81.052327,10.1103/PhysRevA.94.022325,10.1103/PhysRevX.8.021001,10.1103/PhysRevLett.125.040501,10.1103/PhysRevA.104.012431}:
\begin{eqnarray}
|\bar{x}\rangle\otimes|N_x\rangle\equiv|x=\bar{x}+l_xN_x\rangle,
\end{eqnarray}
with $N_x$ being the integer and the modular variable $x\in(0,l_x]$, as illustrated in Fig.~\ref{fig2}. While the choice of $l_x$ is in principle arbitrary, there exists an optimal value to capture the spin properties of the optical spins in Eq.~(\ref{coherentspin}), e.g., the separation of the wave packets in Fig.~\ref{fig2}. This optimal value is usually
\begin{eqnarray} \label{Eq:optimal_ell}
l_{x}^{\rm opt}=2\sqrt{2}\alpha
\end{eqnarray}
for real $\alpha$.

A continuous variable system can now be mapped onto a discrete system by tracing out the modular part:
\begin{eqnarray}
\rho_{\rm discrete}=\int_0^{l_x}d{\bar{x}}\langle\bar{x}|\rho|\bar{x}\rangle.
\end{eqnarray}
The integer variable $N_x$ can be further regrouped into a chain of cells with two internal states each,
\begin{eqnarray}
|m\rangle_{\rm cell}\otimes|n\rangle_{\rm es}\equiv |N_x=2m+n+1\rangle,
\end{eqnarray}
with $m$ being the integer cell index and the internal state label $n=0,1$. By tracing out the cell space $|m\rangle_{\rm cell}$, we then obtain an effective spin state:
\begin{eqnarray}\label{effectivespinreduce}
\rho_{\rm es}&=&\sum_m\langle m|_{\rm cell}\rho_{\rm discrete}|m\rangle_{\rm cell}\nonumber\\
             &=&\sum_m\int_0^{l_x}d{\bar{x}}\langle m|_{\rm cell}\otimes\langle \bar{x}|\rho|\bar{x}\rangle\otimes|m\rangle_{\rm cell} .
\end{eqnarray}
This mapping hence allows us to reinterpret arbitrary continuous-variable states $\rho$ as effective spin states $\rho_{\rm{es}}$, including the steady states of DOPOs. In particular, with the choice (\ref{Eq:optimal_ell}), the mapping (\ref{coherentspin}) is recovered. Finally, it is straightforward to verify that
\begin{eqnarray}
\bar{\sigma_{z}}|\bar{x}\rangle\otimes|m\rangle_{\rm cell}\otimes|n\rangle_{\rm es}&=&-1^{n}|\bar{x}\rangle\otimes|m\rangle_{\rm cell}\otimes|n\rangle_{\rm es},\nonumber\\
\bar{\sigma_{x}}|\bar{x}\rangle\otimes|m\rangle_{\rm cell}\otimes|n\rangle_{\rm es}&=&|\bar{x}\rangle\otimes|m\rangle_{\rm cell}\otimes|1-n\rangle_{\rm es},\nonumber\\
\bar{\sigma_{y}}|\bar{x}\rangle\otimes|m\rangle_{\rm cell}\otimes|n\rangle_{\rm es}&=&-1^ni|\bar{x}\rangle\otimes|m\rangle_{\rm cell}\otimes|1-n\rangle_{\rm es}.\nonumber\\
\end{eqnarray}
This shows that the mapping (\ref{effectivespinreduce}) indeed satisfies the condition (\ref{espindecomposed}), where the ancillary space is composed of the modular space and the cell space,
\begin{eqnarray}
I_{\rm anci}\equiv\sum_m\int_0^{l_x}d{\bar{x}}|m\rangle_{\rm cell}\otimes|\bar{x}\rangle\langle \bar{x}|\otimes\langle m|_{\rm cell}.\nonumber
\end{eqnarray}

\section{Numerical simulations for coherent-cluster-state generation}
In this section, we employ numerical simulations to investigate the generation of coherent cluster states in DOPOs. In the following, we focus on the simplest case of two modes, which is expected to provide lower bounds on the parameter requirements. We also set the two-photon loss rate to be $\Gamma_{\rm d}=1$.
\subsection{Basic picture without single-photon loss}
\begin{figure}[t]
\center
\includegraphics[width=2.8in]{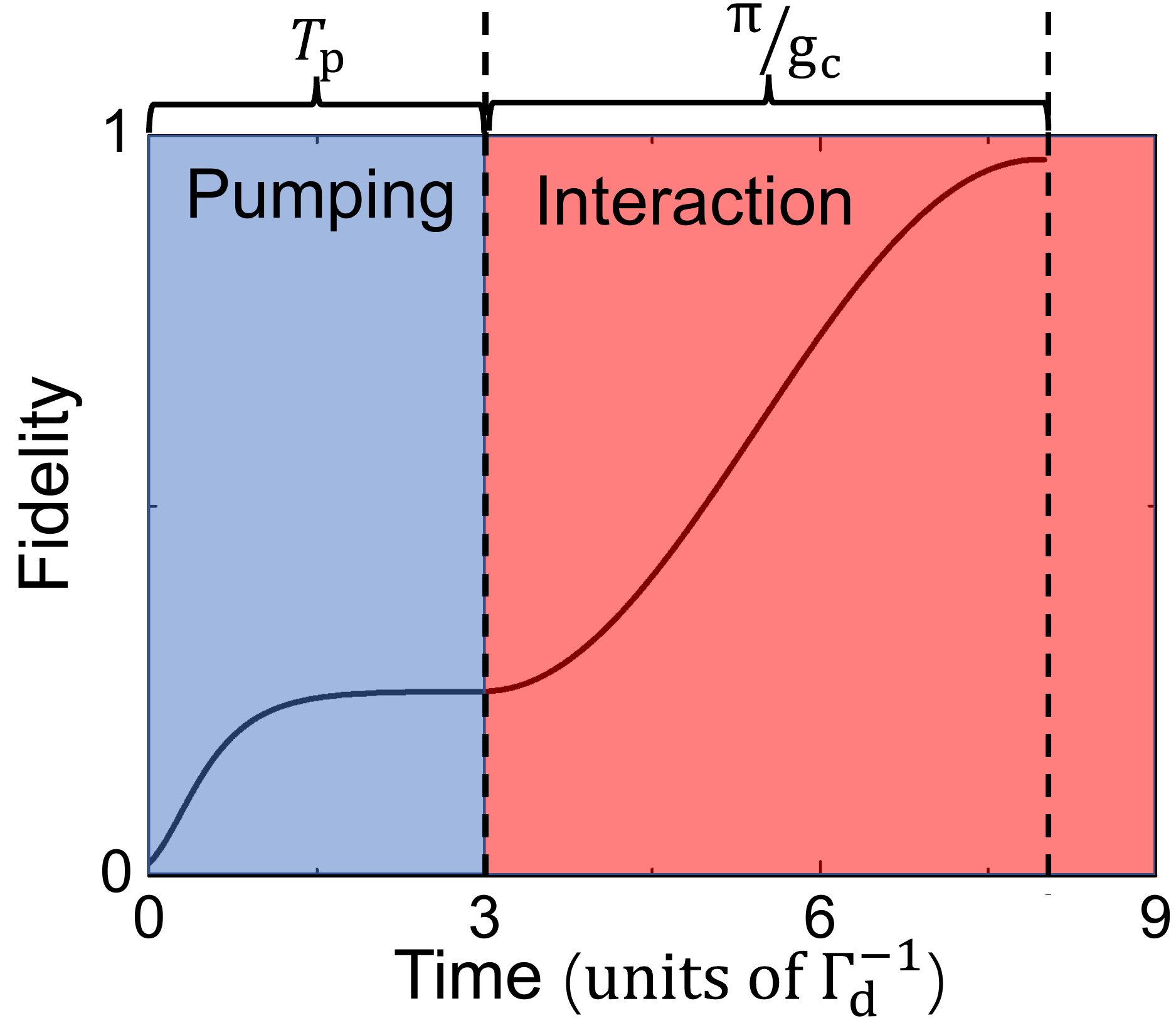}
\caption{Sequential coherent-cluster-state generation, divided into a pumping stage and an interaction stage ($N=2$). Depicted is the fidelity with respect to the ideal coherent cluster state in Eq.~(\ref{coherentcluster2m}) as a function of time, in the absence of single-photon loss.}\label{fig3}
\end{figure}
We first consider the case without single-photon loss, which provides insight into the mechanism of generating coherent cluster states. The system is described by two DOPOs initially in the vacuum state. The two-photon pumps (\ref{twophotonpump}) and the two-photon losses (\ref{twophotoloss}) in each DOPO drive the total system into a separable two-mode cat state (\ref{catplus}). After a time period $T_{\rm p}=3\Gamma_{\rm d}^{-1}$, we further introduce the coherent Ising interaction (\ref{coherentising}) to generate the targeted coherent cluster state (\ref{coherentcluster2m}). Figure~\ref{fig3} exemplifies such a process.

At the initial time in Fig.~\ref{fig3}, the fidelity is very low because the vacuum state is almost orthogonal to the coherent cluster state. The system begins to approach the target state during the pump part and reaches a steady fidelity around $0.25$, which is about the fidelity between the direct product of two cat states and a coherent cluster state. When the coherent Ising interaction terms (\ref{coherentising}) are turned on, the fidelity continues to increase and finally reaches a value close to unity. Figure~\ref{fig3} thus shows that coherent cluster states can be generated based on the coherent Ising interaction, with the latter being realizable with only beam-splitter interactions and classical pumps only. Note that the interacting part here is not sufficiently slow to achieve the fidelity $1$.

\begin{figure}[t]
\center
\includegraphics[width=3.2in]{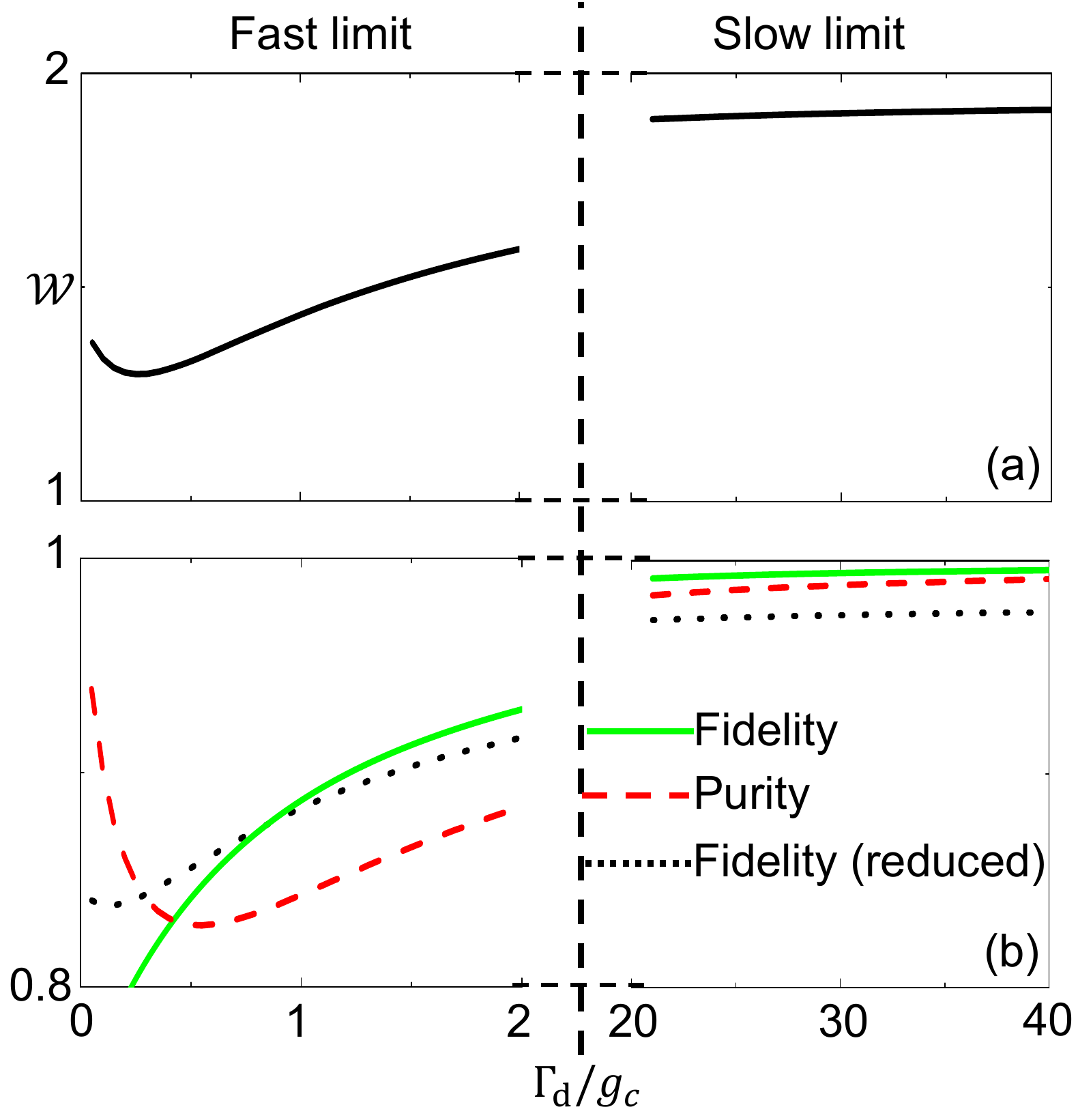}
\caption{Influence of the effective Ising interaction strength (which controls the generation speed) $g_{\rm c}$ on the coherent-cluster-state generation $(N=2)$, as reflected by the fidelity, the purity, and the cluster-state entanglement witness $\mathcal{W}$. Here the duration of the pumping part is $3\Gamma_{\rm d}^{-1}$, the two-photon pumping intensity is $S=-1$, and the duration of the Ising interaction part is $t=\pi /g_{\rm c}$. }\label{fig4}
\end{figure}

The adiabatic requirement $g_{\rm c}\ll\Gamma_{\rm d}$ for the effective Ising interaction~(\ref{coherentising}) has a strong effect on generating coherent cluster states. To see this, we consider in Fig.~\ref{fig4} the influence of $g_{\rm c}$ on the values of different qualifiers for quantum states. The duration of the pumping part is fixed to be $3\Gamma_{\rm d}^{-1}$, which is large enough according to Fig.~{\ref{fig3}}. Note that the interaction strength $g_{\rm c}$ decides the speed of the cluster-state generation, because the interacting time is $t=\pi/g_{\rm c}$.

In Fig.~\ref{fig4}(a), we can see that entanglement can always be detected for different choices of $\Gamma_{\rm d}/g_{\rm c}$, when a comparatively good separable cat state has been generated in the pumping part. The value of the qualifier decreases with growing $\Gamma_{\rm d}/g_{\rm c}$ for large $g_{\rm c}$, which seems to contradict the adiabatic requirement. However, note that the change in the entanglement qualifier is not directly related to the quality of the cluster state unless the threshold $(\mathcal{W}=1)$ is crossed. In addition, the qualifier $\mathcal{W}$ does not reach the ideal value of $2$ in the limit of vanishing $g_{\rm c}$. This can be traced back to two different issues. First, the ideal coherent cluster state (\ref{coherentcluster2m}) is not a perfect cluster state due to the overlap between $|\alpha\rangle$ and $|-\alpha\rangle$. Second, the approximate separable cat state generated in the pumping part is not sufficiently pure.

To better clarify these points, we show in Fig.~\ref{fig4}(b) the corresponding purity and fidelity evolution. The fidelity with respect to an ideal coherent cluster state (green solid curve) always increases with decreasing $g_{\rm c}$. This result is consistent with the adiabatic requirement. The purity of the state (red dashed curve) can provide more insight into this adiabatic requirement. It is not surprising to see a nearly pure state in the slow limit because in this limit the effective Ising interaction in Eq.~(\ref{coherentising}) is equivalent to an exact Ising interaction Hamiltonian. When $g_{\rm c}$ is very large, the purity is also high. In this short-time limit, the two-photon loss has little influence on the system state. However, the purity drops significantly for moderate $g_{\rm c}$. In this regime, the nonadiabatic effects drive the system out of the dark space generated by the two-photon loss and the two-photon pump, so the purity is reduced by the two-photon loss. Although a high purity does not ensure a large qualifier $\mathcal{W}$ or higher entanglement, it may provide more insight into the counter-intuitive results displayed in Fig.~\ref{fig4}(a).

Next, we express the state within the effective spin space according to Eq.~(\ref{effectivespinreduce}) and calculate the fidelity of this reduced state with respect to the reduced state of an ideal coherent cluster state (\ref{coherentcluster2m}) [black dotted curve in Fig.~\ref{fig4}(b)]. This reduced fidelity exhibits the same trend dependence versus $g_{\rm c}$ as the entanglement qualifier $\mathcal{W}$. Although the state generated with a large $g_{\rm c}$ differs from the ideal coherent cluster state (green solid curve), these two states become similar in the reduced space (black dotted curve). Accordingly, the reduced space appears robust to specific imperfections as the effective spin does not depend on the details of the continuous variables in the modular position space $|\bar{x}\rangle$. In the slow limit, although the states generated are very close to an ideal coherent cluster state, the difference is enlarged in the reduced space.

From Fig.~{\ref{fig4}}, we can conclude that the intuitive mapping between the coherent states and the spins is in general correct. However, an exact mapping, as described by Eq.~(\ref{effectivespinreduce}), can in general provide more precise results.

\begin{figure}[t]
\center
\includegraphics[width=3.4in]{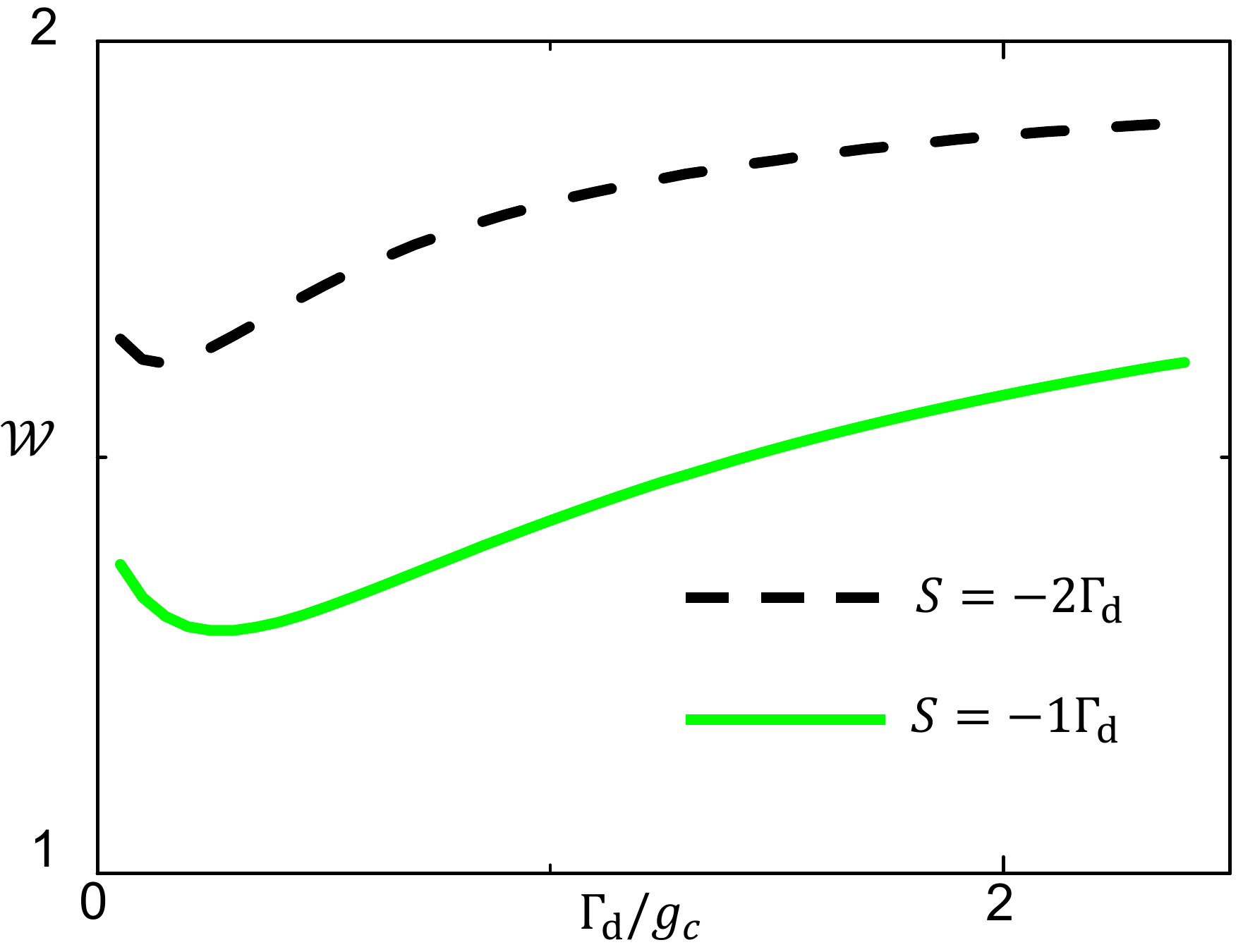}
\caption{Influence of the two-photon pump intensity $S$ on the effective Ising interaction stage of the coherent-cluster-state generation $(N=2)$. The initial state is a two-mode separable cat state with $\alpha=i\sqrt{2S/\Gamma_{\rm d}}$.}\label{fig5}
\end{figure}

In addition to the coupling strength $g_{\rm c}$, the amplitude of the coherent state $|\alpha|$ can also influence the effective rotation. Although a larger-scale $|\alpha|$ can make the state more vulnerable to single-photon loss, it may also improve the cluster-state generation. The adiabatic approximation can be improved due to the coefficient $g_{\rm c}/\alpha$, and the precision of the coherent-to-spin mapping is influenced by the overlap $|\langle-\alpha|\alpha\rangle|$. We demonstrate this influence with the entanglement qualifiers $\mathcal{W}$ for different pumping intensities $S$ (recall that $\alpha=i\sqrt{2S/\Gamma_{\rm d}}$) in Fig.~\ref{fig5}.

As the pumping intensity can also influence the quality of the cat state generated for a given time~\cite{10.1103/PhysRevLett.127.093602}, we do not consider the pumping part and set the initial state to be a two-mode separable cat state (\ref{catplus}) with $\alpha=i\sqrt{2S/\Gamma_{\rm d}}$ and $N=2$. Figure~\ref{fig5} shows that the qualifier $\mathcal{W}$ significantly increases for larger $|S|$. We stress that the value of $\mathcal{W}$ does not quantify the amount of entanglement but merely ensures entanglement when surpassing the threshold of $1$. However, a larger $\mathcal{W}$ can be more robust to single-photon loss if the influence of the single-photon loss rate $\Gamma_{\rm s}$ on $\mathcal{W}$ is continuous. Therefore, there should be an optimal value of $S$ for which the single-photon loss is included.
\begin{figure}[t]
\center
\includegraphics[width=3.4in]{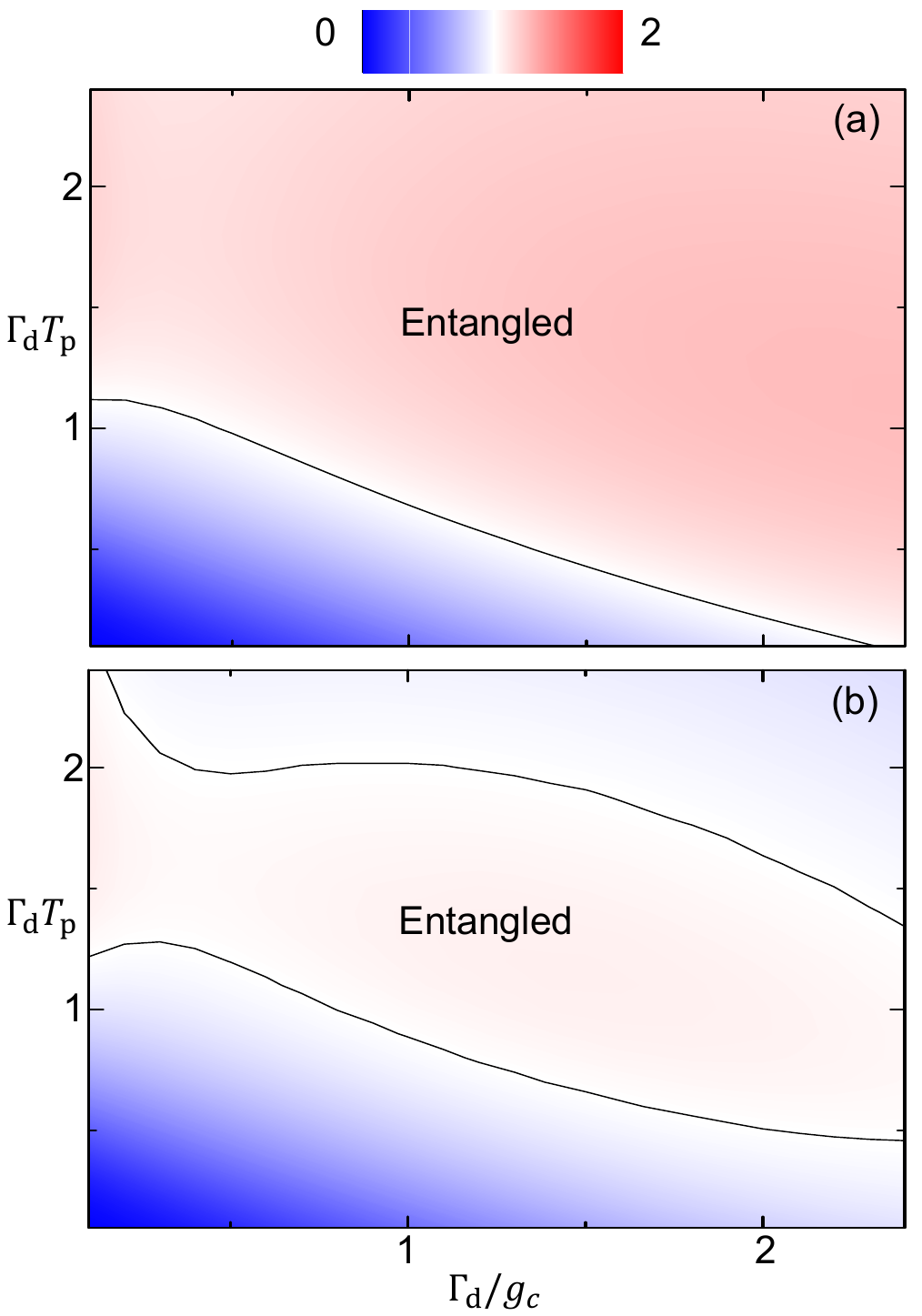}
\caption{Influence of the duration of the pumping stage and the duration of the interaction stage on the coherent-cluster- state generation with the presence of single-photon loss. Depicted is the cluster-state entanglement witness, where the black solid curves indicate the entanglement detectable threshold. The initial state is a vacuum state; the pumping strength is $S=-1\Gamma_{\rm d}$. The single-photon loss rate is $\Gamma_{\rm s}=0.01\Gamma_{\rm d}$ for (a) and $\Gamma_{\rm s}=0.02\Gamma_{\rm d}$ for (b). }\label{fig6}
\end{figure}
\subsection{The optimal parameters with single-photon loss}

Cat-like states are usually vulnerable to single-photon loss~\cite{10.1103/PhysRevA.81.042311,10.1103/PhysRevA.100.012124,10.1103/PhysRevResearch.2.043387,10.1103/PhysRevLett.126.023602,10.1103/PhysRevLett.127.093602}; similarly, the single-photon loss represents the main detrimental effect in coherent-cluster-state generation. We describe the effects of the single-photon loss with the following Lindblad terms:
\begin{eqnarray}\label{singlephotoloss}
\mathcal{L}_{\rm s}(\rho)&=&\sum_{n=1}^2\frac{\Gamma_{\rm s}}{2}[2a_n\rho(t)a_n^{\dag}-a_n^{\dag}a_n\rho(t)-\rho(t)a_n^{\dag}a_n],\nonumber\\
\end{eqnarray}
where we assume the same single-photon loss rate $\Gamma_{\rm s}$ for the two modes in Fig.~\ref{fig6}. We numerically simulate the process described in Fig.~\ref{fig3} and show the value of the entanglement qualifier $\mathcal{W}$ at the end of the evolution.

In Fig.~\ref{fig6}(a), we consider a weak single-photon loss ($\Gamma_{\rm s}=0.01\Gamma_{\rm d}$). In this case, the condition for entanglement generation is quite simple. When the pumping time $T_{\rm p}$ is short, we need a longer interaction time $\pi/g_{\rm c}$. For a short interaction time $\pi/g_{\rm c}$, a longer pumping time $T_{\rm p}$ is required. Extending the duration for either the pumping part or the interacting part does not prevent entanglement generation. This result is similar to the ideal case: A higher-quality separable cat state can be generated in the pumping stage with a longer time, and weak interaction terms better satisfy the adiabatic condition.

In Fig.~\ref{fig6}(b), the single-photon loss rate is $\Gamma_{\rm }=0.02\Gamma_{\rm d}$, which results in a different trend for entanglement generation. There are two boundary curves in Fig.~\ref{fig6}(b). The lower boundary is similar to the boundary in Fig.~\ref{fig6}(a) except for the large-$g_{\rm c}$ limit, where increasing the duration of the interaction part can prevent entanglement generation. The upper boundary is not shown in Fig.~\ref{fig6}(a) but should also exist for larger $T_{\rm p}$ and $g_{\rm c}^{-1}$. Note that the single-photon loss usually prohibits entanglement in the long-time limit. For most of this upper boundary, entanglement generation can be prevented by increasing either $\pi/g_{\rm c}$ or $T_{\rm p}$. Therefore, the lower boundary is related to the breakdown of the steady-state condition and the adiabatic condition, so that parameters near this boundary prefer a slower evolution; the upper boundary, in contrast, is a consequence of the single-photon loss, so that a faster evolution can improve the coherent-cluster-state generation. The opposite trends at the two boundaries at large values of $g_{\rm c}$ can be seen as the result of the nonadiabatic effects and the two-photon loss, indicated in Fig.~\ref{fig4}(a).
\begin{figure}[t]
\center
\includegraphics[width=3.4in]{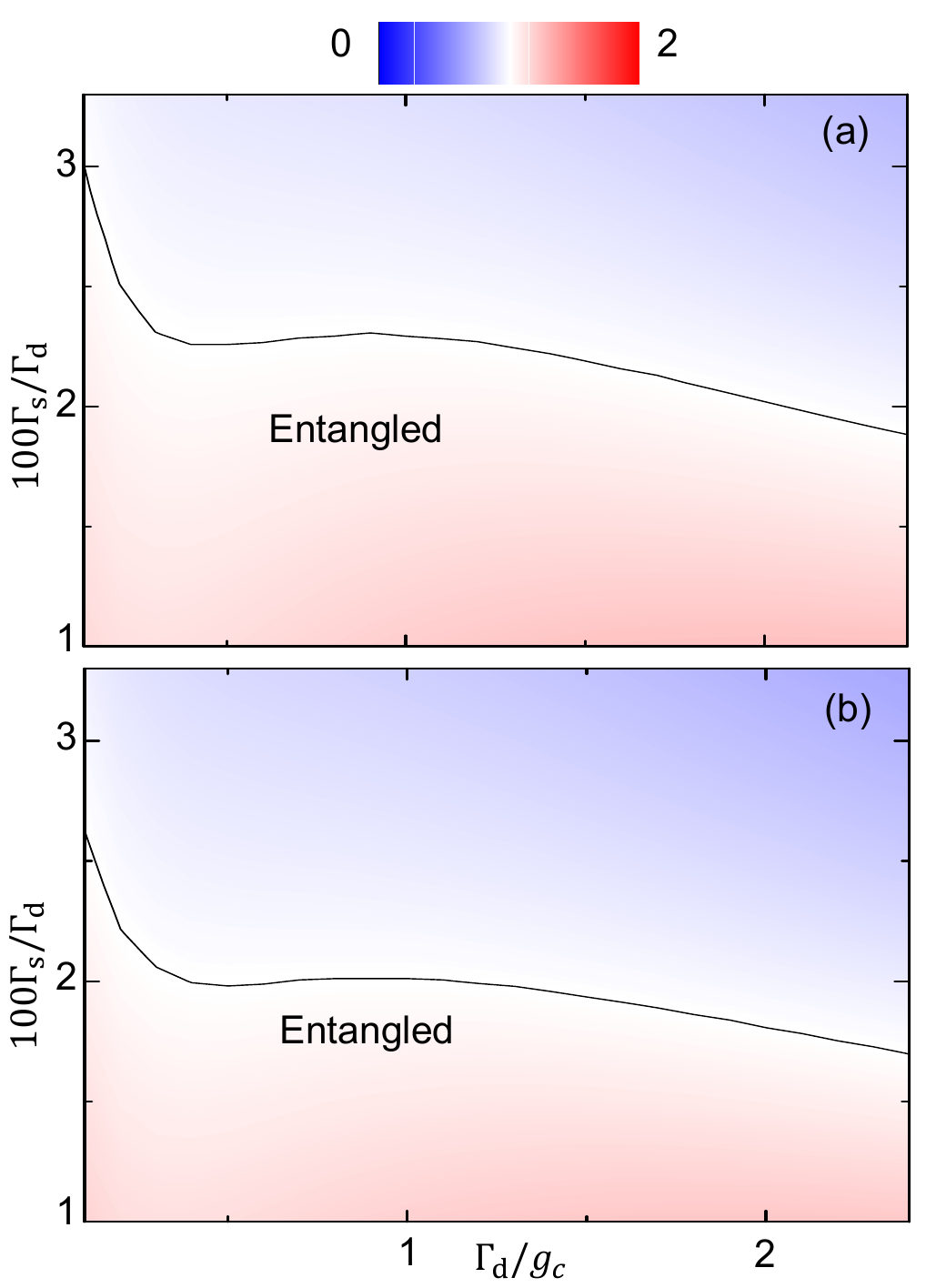}
\caption{Influence of the single-photon rate $\Gamma_{\rm d}$ and the duration of the interaction part on the generation of coherent cluster states in the presence of single-photon loss. The initial state is a vacuum state; the pumping strength is $S=-1\Gamma_{\rm d}$. The pumping duration is $T_{\rm p}=1.6\Gamma^{-1}_{\rm d}$ for (a) and $T_{\rm p}=2\Gamma^{-1}_{\rm d}$ for (b). }\label{fig7}
\end{figure}

In Fig.~\ref{fig6}, we can see that the pumping times around $\Gamma_{\rm d}T_{\rm p}\sim1.5$ are less affected by the single-photon loss. Therefore, we choose two different values of $\Gamma_{\rm d}T_{\rm p}$, and show the influence of the single-photon loss rate $\Gamma_{\rm s}$ and the interaction strength $\Gamma_{\rm d}/g_{\rm c}$ in Fig.~\ref{fig7}. In Figs.~\ref{fig7}(a) and~\ref{fig7}(b), the entanglement boundaries have similar shapes, although entanglement is generated with a higher single-photon loss rate in Fig.~\ref{fig7}(a). Therefore, there can be an optimal value of $T_{\rm p}$ while the mechanism for cluster-state generation remains the same. However, the value of $g_{\rm c}$ may influence the mechanism for cluster-state generation, as both boundaries in Fig.~\ref{fig7} can be divided into two parts. When $\Gamma_{\rm d}/g_{\rm c}$ is smaller than $0.5$, we see fast drops in the boundary curves. After this value, the curves change less significantly with $\Gamma_{\rm d}/g_{\rm c}$. Similar changes around $\Gamma_{\rm d}/g_{\rm c}\sim0.5$ are observed in Figs.~\ref{fig4}, \ref{fig5}, and \ref{fig6}. Note that the (approximate) coherent cluster states generated in the fast limit are more robust to single-photon loss, although they exhibit a reduced quality in the ideal limit of vanishing single-photon loss. The highest single-photon rate in Fig.~\ref{fig7}, under which entanglement can be generated, is about $0.03\Gamma_{\rm d}$. This value is lower than the one for the entangled-cat-state generation~\cite{10.1103/PhysRevA.104.013715}, but at the same order of magnitude. Therefore, the coherent-cluster-state generation, while being more vulnerable to single-photon loss compared to the entangled-cat-state generation, imposes a similar requirement on the control of the single-photon loss.
\begin{figure}[t]
\center
\includegraphics[width=3.4in]{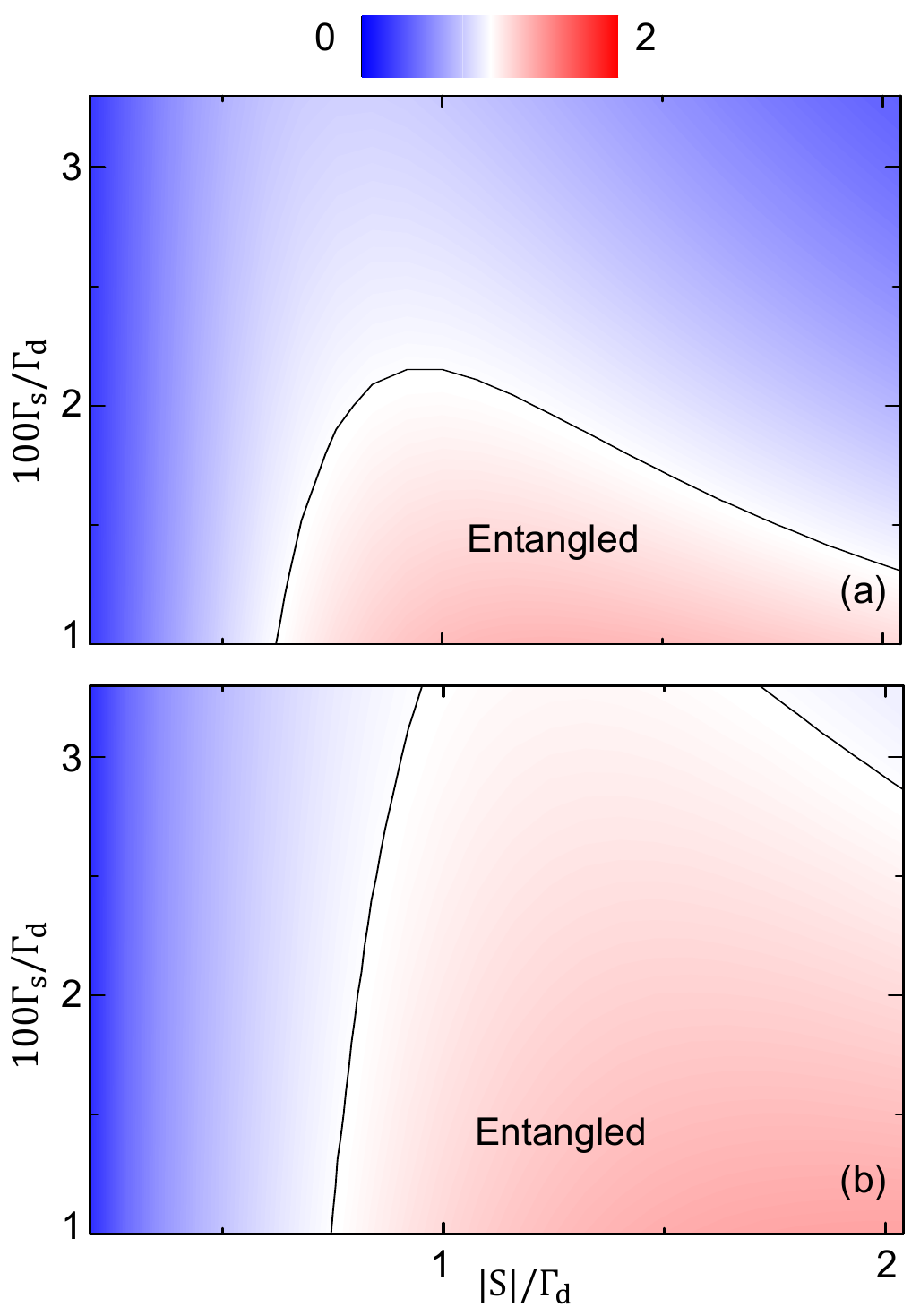}
\caption{Influence of the single-photon loss rate $\Gamma_{\rm d}$ and the two-photon pumping intensity $S$ on the generation of coherent cluster states in the fast limit and the slow limit. The initial state is a vacuum state; the pumping time is $T_{\rm p}=1.5\Gamma^{-1}_{\rm d}$. (a) The effective Ising interacting is moderate, $g_{\rm c}=\Gamma_{\rm d}/1.5$. (b) The effective Ising interacting is strong, $g_{\rm c}=20\Gamma_{\rm d}$. }\label{fig8}
\end{figure}

As the cluster-state generation is also affected by the two-photon pumping intensity $S$, as shown in Fig.~\ref{fig5}, we finally consider the relation between $S$ and the highest single-photon loss rate $\Gamma_{\rm s}$ under which entanglement is detectable. The pumping duration is fixed to be $1.5\Gamma_{\rm d}^{-1}$, which can generate detectable entanglement for a wide range of $g_{\rm c}$ in the presence of single-photon loss according to Fig.~\ref{fig6}. In Figs.~\ref{fig4}-\ref{fig7}, the results change abruptly near $g_{\rm c}\sim2\Gamma_{\rm d}$, so we consider two different values of the effective Ising interaction strength, specifically, $g_{\rm c}=\Gamma_{\rm d}/1.5$ and $g_{\rm c}=20\Gamma_{\rm d}$.

Figure~\ref{fig8} indicates the optimal values for the two-photon pumping intensity in both cases, as expected in the previous subsection. When the pumping intensity $S$ is below the optimal value, the entanglement generation is more sensitive to $S$. For a moderate $g_{\rm c}=\Gamma_{\rm d}/1.5$, the optimal value of $S$ is about $\Gamma_{\rm d}$. The highest single-photon loss for entanglement generation is about $0.02\Gamma_{\rm d}$, which is not significantly different from the results in Figs.~\ref{fig6} and \ref{fig7}. In addition to the higher tolerance to single-photon loss shown for large values of $g_{\rm c}$ in Fig.~\ref{fig8}, we can also find a shift of the optimal $S$ to a larger value in Fig.~\ref{fig8}(b). This shift is the result of less single-photon loss due to a shorter interaction duration $\pi/g_{\rm c}$. Note that the entanglement qualifier $\mathcal{W}$ prefers large pumping intensities $|S|$ in the ideal case as shown in Fig.~\ref{fig5}.

In this section, we considered the coherent-cluster-state generation in the presence of single-photon loss. We find that, to achieve a higher tolerance to the single-photon loss, there are optimal values of the pumping time $T_{\rm p}$ and the pumping intensity $S$, which are both between $\Gamma_{\rm d}$ and $2\Gamma_{\rm d}$. A fast effective Ising interaction is always preferable in the presence of single-photon loss, in spite of the detrimental nonadiabatic effects.

\subsection{Effects of nonequilibrium pumping}
\begin{figure}[t]
\center
\includegraphics[width=3.4in]{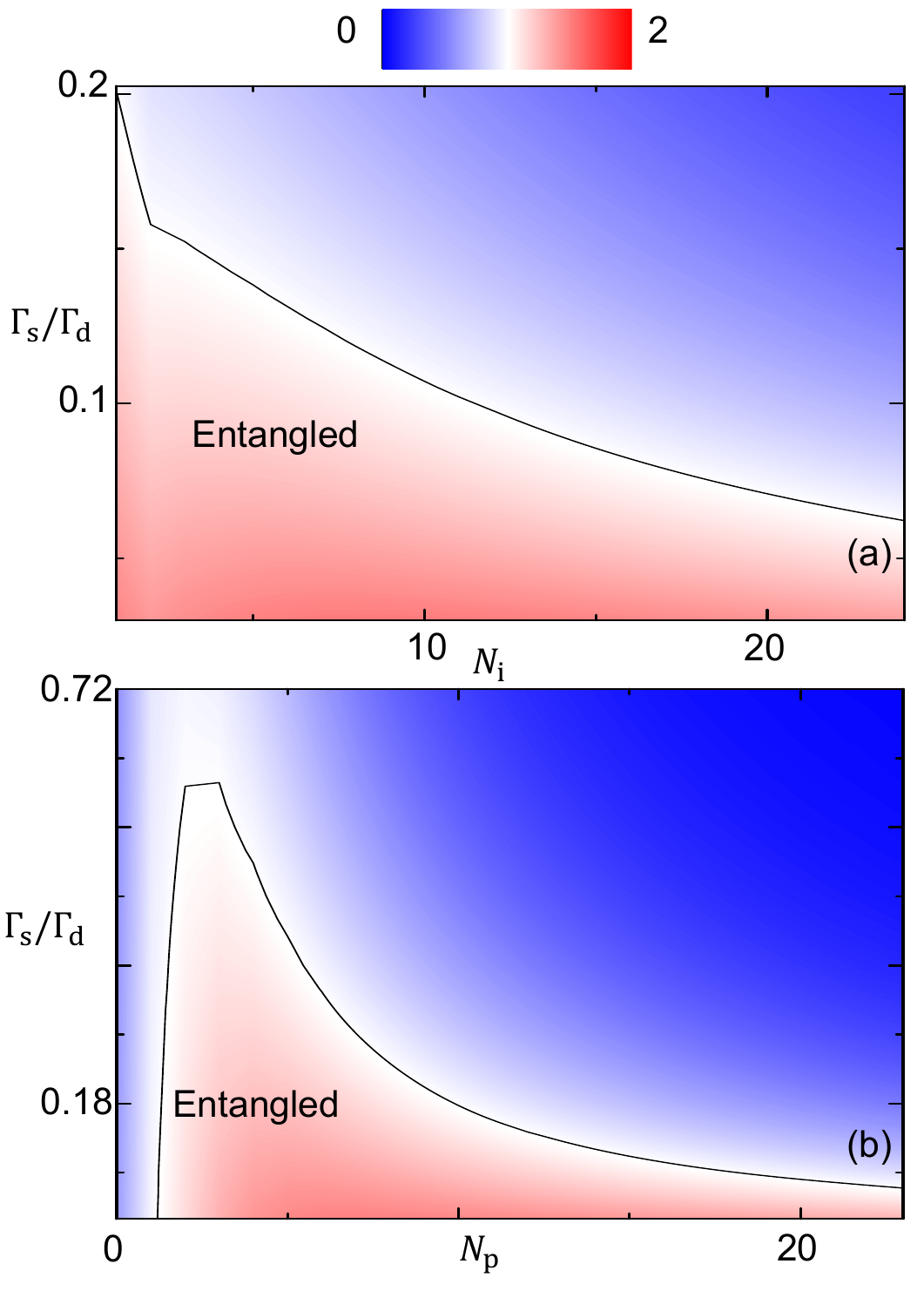}
\caption{Performance of the coherent-cluster-state generation with nonequilibrium pumps in the presence of single-photon loss. (a) The number of pumping cycles is $N_{\rm p}=9$. (b) Results corresponding to $N_{\rm i}=1$ in (a) with different numbers of pumping cycles $N_{\rm p}$.}\label{fig9}
\end{figure}
It has been shown that nonequilibrium pump fields can be used in DOPO systems to improve the cat-state generation~\cite{10.1103/PhysRevA.106.023714}. We now demonstrate that a similar improvement can be achieved for the generation of coherent cluster states.

The effects of the equilibrium model~(\ref{twophotonpump}) and (\ref{twophotoloss}) can also be realized by a cyclic model:
\begin{eqnarray}
\rho_{n+1}&=&{\rm Tr}_{b}\left\{e^{-iH^{\rm I}_{\rm nl}t_{\rm nl}}\rho_{n}\otimes |\alpha_{\rm p}\rangle_{\rm c}\langle\alpha_{\rm p}|_{\rm c}e^{iH^{\rm I}_{\rm nl}t_{\rm nl}}\right\},\nonumber
\end{eqnarray}
with,
\begin{eqnarray}\label{intnonlinearhamiltonian}
H^{\rm I}_{\rm nl}&=&\sum_{n=1}^2g_{\rm nl}[b_n^{\dag}a_n^2+b_n(a_n^{\dag})^2] , \nonumber\\
|\alpha_{p}\rangle_{\rm c}&=&|\alpha_{p}\rangle\otimes|\alpha_{p}\rangle.
\end{eqnarray}
Here $g_{\rm nl}$ is the nonlinear interaction strength, $b_{n}$ is the annihilation operator of the pump mode of the $n$th DOPO mode, and $|\alpha_{p}\rangle_{\rm c}$ is the direct product of two pump-mode states. The loss can be described by additional Lindblad terms in Eq.~(\ref{intnonlinearhamiltonian}):
\begin{eqnarray}\label{cyclicwithloss}
\frac{\partial {\rho}_n(t)}{\partial t}&=&-i[H^{\rm I}_{\rm nl},\rho_n(t)]+\mathcal{L}_{\rm s}(\rho_n(t))+\mathcal{L}_{\rm p}(\rho_n(t)),\nonumber\\
\mathcal{L}_{\rm s}(\rho)&=&\sum_{n=1}^2\frac{\Gamma_{\rm s}}{2}[2a_n\rho(t)a_n^{\dag}-a_n^{\dag}a_n\rho(t)-\rho(t)a_n^{\dag}a_n],\nonumber\\
\mathcal{L}_{\rm p}(\rho)&=&\sum_{n=1}^2\frac{\Gamma_{\rm s}}{2}[2b_n\rho(t)b_n^{\dag}-b_n^{\dag}b_n\rho(t)-\rho(t)b_n^{\dag}b_n],\nonumber\\
\end{eqnarray}
with $\rho_n(0)=\rho_{n}\otimes |\alpha_p\rangle_{\rm c}\langle\alpha_p|_{\rm c}$ and $\rho_{n+1}={\rm Tr}_{b}\left\{\rho_n(t_{\rm nl})\right\}$. The amplitudes of the pump modes are chosen to generate steady DOPO amplitudes $\alpha_i=\sqrt{2}$. We set the duration of the nonlinear pumping for each cycle to be $t_{\rm nl}=0.5g_{\rm nl}^{-1}$. Instead of the pumping time $T_{\rm p}$ and the interaction time $\pi/g_{\rm c}$ used in the previous sections, we describe the duration of each part with the number of pumping cycles $N_{\rm p}$ and the number of interaction cycles $N_{\rm i}$. The initial state is assumed to be a vacuum state.

In the pumping part, the system experiences the cyclic dynamics (\ref{cyclicwithloss}) for $N_{\rm p}$ cycles. In the interacting part, the system dynamics is described by cycles composed of two parts. The first part is described by Eq.~(\ref{cyclicwithloss}), which contains single-photon loss. The second part, which lasts for the time period $t=\pi/(g_{\rm c}N_{\rm i})$, consists of the beam-splitter interaction and the classical pumping as shown in Eq.~(\ref{coherentising}). To simplify the problem, we assume $g_{\rm c}$ is sufficiently large that we can ignore the influence of the single-photon loss in this part. Note that the interaction can be adiabatic with a large $g_{\rm c}$ if we have a large $N_{\rm i}$. After $N_{\rm i}$ cycles of interaction dynamics, we calculate the entanglement qualifier $\mathcal{W}$. For an easier comparison with the adiabatic results described in Eqs.~(\ref{twophotonpump}) and (\ref{twophotoloss}), we assume the ratio between the adiabatic two-photon loss rate $\Gamma_{\rm d}$ and the nonlinear coupling intensity $g_{\rm nl}$ is the same as in the experiment~\cite{10.1126/science.aaa2085}, i.e., $g_{\rm nl}=15\Gamma_{\rm d}$. However, note that the nonequilibrium pumping cannot be characterized by an effective two-photon loss rate $\Gamma_{\rm d}$.

In Fig.~\ref{fig9}(a), we set $(N_{\rm p}=9)$, which is sufficiently large to obtain a good two-mode separable cat state according to our previous work~\cite{10.1103/PhysRevA.106.023714}, and consider the influence of the number of interaction cycles $N_{\rm i}$ on the cluster-state entanglement witness. Similar to Fig.~\ref{fig7}, a ``fast" effective Ising interaction is preferable for generating entanglement under high single-photon loss rates $\Gamma_{\rm s}$. We find that the highest tolerated single-photo loss rate is about $7$ times larger by introducing the nonequilibrium pumping method. In Fig.~\ref{fig9}(b), we consider the influence of the number of pumping cycles $N_{\rm p}$. The number of interacting cycles is set to be the optimal value $(N_{\rm i}=1)$ as found in Fig.~\ref{fig9}(a). The tolerance to the single-photon loss can be further enhanced by choosing an optimal $N_{\rm p}$ according to Fig.~\ref{fig9}(b). Note that the curves in Fig.~\ref{fig9} are not smooth because both $N_{\rm p}$ and $N_{\rm i}$ can take only integer values.

\subsection{Discussion of the required parameters}
Based on our numerical results, it can be concluded that a minimum nonlinear coupling strength of approximately $g_{\rm nl}\approx20\Gamma_{\rm s}$ is required for generating coherent cluster states. Achieving such a nonlinearity-to-loss ratio is challenging in current CIMs and DOPO systems based on periodically poled lithium niobate (PPLN) waveguides. Considering the achievable parameters in PPLN, the highest nonlinear coupling strength $g_{\rm nl}$ ranges from $0.01\Gamma_{\rm s}$ to $0.1\Gamma_{\rm s}$~\cite{10.1103/PhysRevA.94.063809,10.1364/OPTICA.5.001438,10.1063/5.0063118}. To attain the required parameter range in PPLN systems, potential solutions include using high-quality PPLN waveguides or employing shorter optical pulses. Another approach is to modify the structure of the existing CIMs, which would not require significant improvements in experimental technology. While current CIMs operate above threshold, it has been successfully demonstrated that optical cluster states can be generated in DOPO systems below threshold~\cite{10.1038/nphoton.2013.287}. Therefore, adjusting the working regime could potentially reduce the parameter requirements. Alternatively, Josephson parametric oscillators~\cite{10.1038/ncomms5480,10.1103/PhysRevX.9.021049}, known for their high nonlinearity-to-loss ratio, could be utilized to construct CIMs in the microwave regime.

\section{Conclusions}

We investigated the generation of coherent cluster states in DOPO networks with effective optical Ising interactions. DOPO networks can produce coherent cluster states based on the generation of separable cat states, beam-splitter interactions, and classical pumping. These coherent cluster states can be mapped to an effective spin Hilbert space formed by modular variables. In this effective spin Hilbert space, we can apply the entanglement criteria designed for spin systems to assess the quality of the state generation in the presence of detrimental effects, e.g., single-photon loss, overlap between coherent states, or nonadiabatic effects.

As a paradigmatic example, we considered the case of two modes, which can be expected to deliver lower bounds on the parameter requirements. We applied the respective stabilizer entanglement criterion to explore the parameter regime that supports the coherent-cluster-state generation. Our results indicate that the single-photon loss acceptable for cluster-state generation must remain below about one third of the single-photon loss that is acceptable for the entangled-cat-state generation. In addition, we found that a nonequilibrium pump can significantly raise the threshold for the coherent-cluster-state generation, with tolerable single-photon loss rates that are about one order of magnitude larger.

This work may help us to explore the quantum properties of, and possibly to realize one-way quantum computation in, DOPO networks, such as coherent Ising machines. Moreover, we hope that our analysis can contribute to elucidating the properties of coherent-state coding spaces.

\begin{acknowledgments}
J.Q.Y. is partially supported by the National Key Research and Development Program of China (Grant No.~2022YFA1405200) and the National Natural Science Foundation of China (NSFC) (Grant No.~92265202 and No.~11934010). F.N. is supported in part by: Nippon Telegraph and Telephone Corporation (NTT) Research, the Japan Science and Technology Agency (JST) [via the Quantum Leap Flagship Program (Q-LEAP), and the Moonshot R{\&}D Grant Number JPMJMS2061], the Asian Office of Aerospace Research and Development (AOARD) (via Grant No. FA2386-20-1-4069), and the Foundational Questions Institute Fund (FQXi) via Grant No. FQXi-IAF19-06.
\end{acknowledgments}


%
\bibliography{totalreference}

\end{document}